\documentclass{Interspeech2024}
\usepackage{multirow}
\usepackage[table,xcdraw]{xcolor}

\interspeechcameraready

\title{The X-LANCE Technical Report for Interspeech 2024 \\ Speech Processing Using Discrete Speech Unit Challenge}

\name[affiliation={1}]{Yiwei}{Guo$^{\ast \spadesuit}$}
\name[affiliation={1}]{Chenrun}{Wang$^{\ast \clubsuit}$}
\name[affiliation={1}]{Yifan}{Yang$^{\ast \heartsuit}$}
\name[affiliation={1}]{Hankun}{Wang$^{\clubsuit}$}
\name[affiliation={1}]{Ziyang}{Ma$^{\spadesuit \heartsuit}$}
\name[affiliation={1}]{Chenpeng}{Du$^{\spadesuit \clubsuit}$}
\name[affiliation={2}]{Shuai}{Wang}
\name[affiliation={3}]{Hanzheng}{Li}
\name[affiliation={3}]{Xu}{Li}
\name[affiliation={3}]{Shuai}{Fan}
\name[affiliation={3}]{Hui}{Zhang}
\name[affiliation={1}]{Xie}{Chen}
\name[affiliation={1,3}]{Kai}{Yu}

\address{
  $^1$MoE Key Lab of Artificial Intelligence, X-LANCE Lab, 
  Shanghai Jiao Tong University, China\\
  $^2$Shenzhen Research Institute of Big Data, China
  $^3$AISpeech Ltd, China}
\email{\{chenxie95, kai.yu\}@sjtu.edu.cn}


\begin{document}

\maketitle

\renewcommand{\thefootnote}{\fnsymbol{footnote}}
\footnotetext{$^\ast$Equal contribution}
\footnotetext{$^\spadesuit$Text-to-speech (TTS) track}
\footnotetext{$^\clubsuit$Singing voice synthesis (SVS) track}
\footnotetext{$^\heartsuit$Automatic speech recognition (ASR) track}
\renewcommand{\thefootnote}{\arabic{footnote}}

\begin{abstract}
    Discrete speech tokens have been more and more popular in multiple speech processing fields, including automatic speech recognition (ASR), text-to-speech (TTS) and singing voice synthesis (SVS). 
    In this paper, we describe the systems developed by the SJTU X-LANCE group for the TTS (acoustic + vocoder), SVS, and ASR tracks in the Interspeech 2024 Speech Processing Using Discrete Speech Unit Challenge.
    Notably, we achieved 1st rank on the leaderboard\footnote{Submitted on March 15th 2024 due to a non-technical issue, thus still evaluated by the organizers.}
    in the TTS track both with the whole training set and only 1h training data, with the highest UTMOS score and lowest bitrate among all submissions.
\end{abstract}

\vspace{-0.05in}
\section{TTS Track: The VQTTS System}
The TTS track requires participants to build a discrete token-based TTS on the LJSpeech dataset, with a favorably low bitrate and high naturalness score.
Our best submission used FunCodec~\cite{du2023funcodec} as the discrete tokens and a modified VQTTS~\cite{VQTTS} system as the TTS pipeline. 
\vspace{-0.1in}
\subsection{Discrete Tokens}
In TTS, two kinds of speech discretization are both favored: the semantic tokens and the acoustic tokens~\cite{audiolm,speartts}.
Semantic tokens capture most of the content-relevant information by clustering on self-supervised learning models, while the acoustic tokens aim to reconstruct speech signals as perfectly as possible.
Both two types exhibit their own pros and cons, and have been successfully used in speech synthesis tasks~\cite{audiolm,speartts,valle,UniCATS,zhu2023vec,song2024ella}.

In this TTS track, we considered a semantic token, wav2vec2.0~\cite{wav2vec2}, and an acoustic token, FunCodec~\cite{du2023funcodec}.
Specifically, for wav2vec2.0, we used the official wav2vec2-large-lv60 model\footnote{https://huggingface.co/facebook/wav2vec2-large-lv60} which was pretrained on 60k hours of LibriLight~\cite{librilight}.
We did not use a finetuned version of wav2vec2.0 because the quantizer is only trained in the pretraining stage.
This led to a codebook of 2 groups, each with 320 vocabulary size and 384-dimensional code-vectors in 50Hz frame rate.
In our preliminary study, the discrete codes from wav2vec2.0 contain more prosodic information than vq-wav2vec~\cite{Baevski2020vq-wav2vec} and Kmeans clusters on HuBERT~\cite{hubert}.
The 50Hz frame rate also benefits the reduction of bitrate.
After extracting the 2 groups of codebook indexes, we identified 24686 unique pairs that occurred at least once in the training set, hence the original 2 groups can be transformed to one group of 24686 integer indexes on the provided corpus with a bitrate at about 729bps.
The 2 groups of 320 codes contain $320^2=102400$ possible combinations, so compressing them into one group considerably reduced the bitrate.

For FunCodec, we used an open-sourced checkpoint\footnote{https://huggingface.co/alibaba-damo/audio\_codec-encodec-zh\_en-general-16k-nq32ds640-pytorch} trained on a large in-house corpus containing both English and Chinese data with an total about 25k hours.
It adopts a similar architecture with EnCodec~\cite{encodec}, with the following advantages:
\begin{itemize}
    \item The frame rate is 25Hz, i.e. 640 times downsampling on 16kHz waveform. This reduces the bitrate significantly, since the necessary vocabulary size can be relaxed in a squared sense when the sequence length is halved.
    \item Some techniques are introduced into training, such as an additional magnitude spectrum loss, structured dropout, and codebook learning strategies~\cite{du2023funcodec}. This improves the reconstruction quality of the quantized codes.
\end{itemize}
We only used the first codebook with 1024-size and 128-dimensional code vectors, and trained a customized vocoder to better adapt to the given corpus. 
The bitrate is \textbf{250bps} then, which is the lowest among all submissions in this track.

\subsection{Model Architecture}
\vspace{-0.05in}
\subsubsection{Acoustic Model}
The acoustic model in this challenge inherits the txt2vec acoustic model in VQTTS~\cite{VQTTS}.
The model architecture is visualized in Fig.\ref{fig:tts}. 
The input phoneme sequence is fed into a conformer-based text encoder, and then a phoneme-level prosody controller.
This prosody controller predicts the Kmeans clustering index of phoneme-averaged dynamic prosodic features\footnote{Pitch, probability of voice, energy, and their first and second-order differences.} via an LSTM. 
Then, a length regulator repeats the phone-level sequence to frame-level.
After a conformer decoder, we design another causal decoder that predicts the vector-quantized (VQ) indexes.
Although this is an autoregressive decoder, the decoding timesteps are fixed to be the durations.
In every decoding step, the input to the causal VQ decoder is the corresponding hidden state concatenated with the code vector of the code index from the last step.
In this practice, we favor the transformer decoder structure rather than the LSTM in VQTTS.
The training criterion consists of the cross entropy of VQ prediction, the duration prediction loss and the prosody prediction loss.

\vspace{-.1in}
\subsubsection{Discrete Unit-based Vocoder}
\begin{figure}
    \centering
    \includegraphics[width=0.9\linewidth]{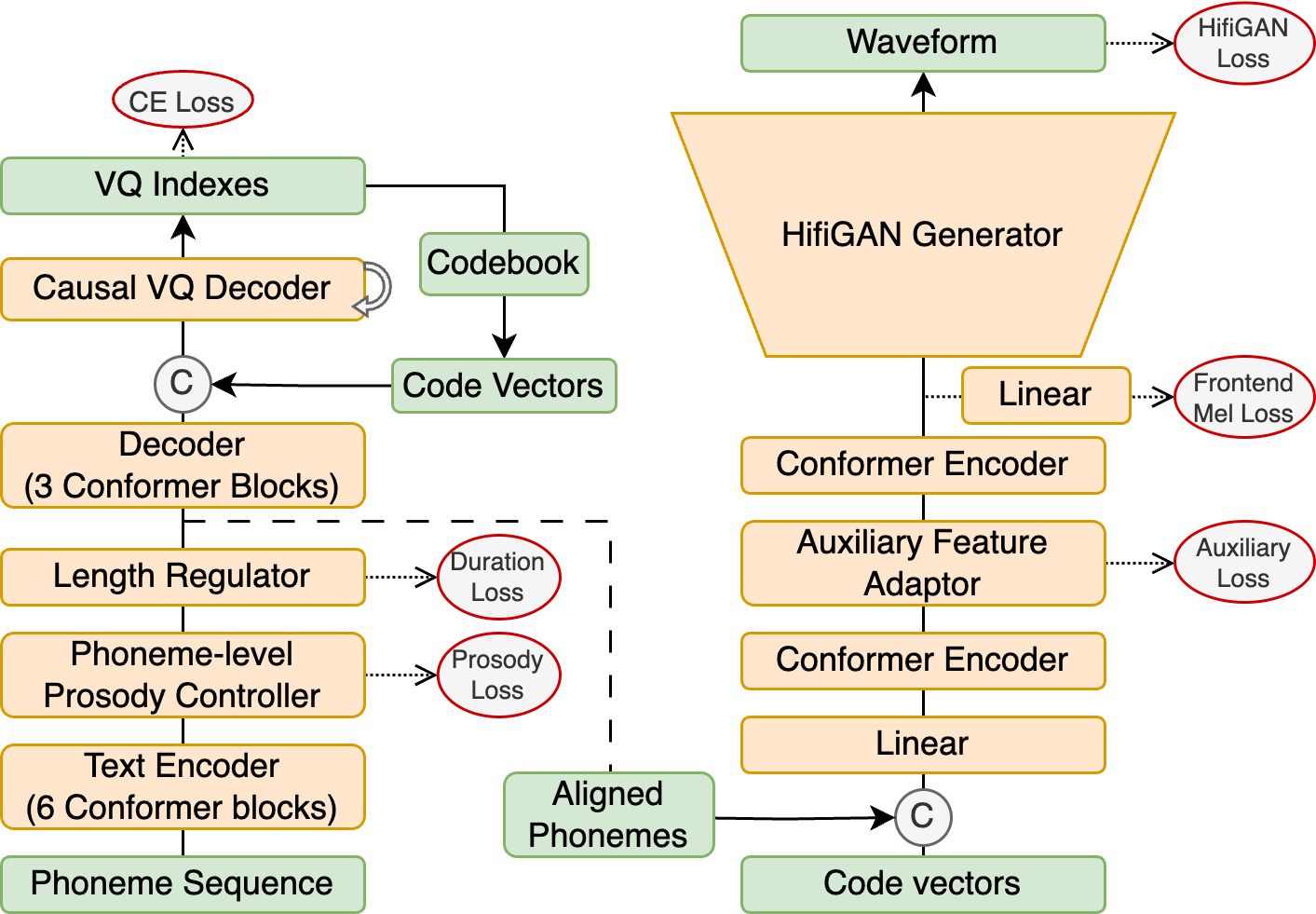}
    \caption{Architecture of the acoustic model (left) and vocoder (right) in the TTS track. ``C" in the circle means to concatenate along dimensions.}
    \label{fig:tts}
\end{figure}
Our vocoder resembles the CTX-vec2wav vocoder introduced in UniCATS~\cite{UniCATS}. We discard the ``context" part since this track only considers single-speaker TTS.
As is shown in Fig.\ref{fig:tts}, the code vectors are first fed into a frontend consisting of conformer encoders and an auxiliary feature adaptor, before entering the generator of HifiGAN~\cite{hifigan}.
The auxiliary feature adaptor predicts the prosody features.
To further mitigate the pronunciation errors brought by speech discretization, we implement a trick that provides the aligned phoneme sequence to the vocoder by embedding it and concatenating it with the code vectors.
As the phonemes are the input to the whole TTS system, this trick more resembles a ``residual connection" and thus does not increase the bitrate.
In inference, the aligned phoneme sequence can be directly obtained from the length regulator in the acoustic model.
This relaxes the burden of the vocoder since phonemes already provide strong information of articulation.

The training criterion of this vocoder includes the HifiGAN loss (i.e. mel loss, feature matching loss and discriminator loss), the frontend mel loss and the auxiliary feature prediction loss.

\subsection{Data Preparation and Training}
\subsubsection{Data Preparation}
The provided dataset is LJSpeech~\cite{ljspeech17}, a single-speaker English corpus with about 24 hours.
We downsampled all speech data to 16kHz.
We used Kaldi~\cite{povey2011kaldi} to extract the 80-dimensional mel-spectrograms and prosodic features in 20ms frame shift for training the vocoder.
These features were cepstral mean-normalized with statistics computed on the training set.

For texts, we obtained optional silence marks by performing alignment in 10ms by Kaldi.
We then needed to have the phoneme duration in 20ms frame shift, but this appeared to be too large for Kaldi since every phoneme must be aligned to at least 3 consecutive frames there.
Hence we resorted to RAD-TTS~\cite{shih2021rad}, a flow-based TTS model with the capability of learning robust alignment paths.
We trained a RAD-TTS model on LibriTTS~\cite{libritts} in 20ms frame shift and then performed inference on the LJSpeech dataset.
However, the FunCodec tokens are in 40ms frame shift, which might still be too large for alignment. 
Hence for FunCodec, we repeated each token for two times to match the 20ms duration both in the acoustic model and vocoder.
This had no effect on the bitrate either, because folding and repeating can both be done inside the models.

\subsubsection{Training}
We trained the acoustic model for 160 epochs using the Adam optimizer at a learning rate of $10^{-3}$ and a decay at $10^{-6}$.
The text encoder and decoder contains 6 and 3 conformer blocks respectively, each with 384 attention dimensions and 2 heads. 
The causal VQ decoder is a transformer decoder with 6 layers, 768 attention dimensions and 12 heads.
The three losses in the acoustic model were equally weighted.

The implementation and training of the vocoder follows that of UniCATS\footnote{https://github.com/X-LANCE/UniCATS-CTX-vec2wav}, only with the difference of input dimensions.

\subsection{Experimental Results}

\begin{table}[]
\centering
\caption{Resynthesis and TTS results in the TTS track. WER is measured by Whisper ``medium.en" version, and ``w/ phn." means to provide aligned phonemes in the vocoder. The official baseline as a bitrate of 448.3bps with UTMOS of 3.73 on the whole training set, and 2.48 on the 1h training set.
}
\label{tab:tts}
\begin{tabular}{@{}lccc@{}}
\toprule
 & \begin{tabular}[c]{@{}c@{}}\textbf{Bitrate}$\downarrow$\\ \footnotesize{(bps)}\end{tabular} & \begin{tabular}[c]{@{}c@{}}\textbf{WER}$\downarrow$ \\ \footnotesize{(Whisper)}\end{tabular} & \textbf{UTMOS}$\uparrow$ \\ \midrule
Ground truth & -  & 2.31 & 4.43$\pm$0.07 \\ \midrule\midrule
\textbf{Resynthesis} &  &  &  \\ \midrule
wav2vec2.0 &  & 2.84 & 4.38$\pm$0.15 \\
wav2vec2.0 w/ phn. & \multirow{-2}{*}{729} & \textbf{2.51} & \textbf{4.44$\pm$0.10} \\
\rowcolor[HTML]{EFEFEF} 
FunCodec w/ phn. & \textbf{250} & 4.14 & 4.43$\pm$0.13 \\ 
\midrule\midrule
\textbf{Resynthesis-1h} &  &  &  \\ \midrule
wav2vec2.0 w/ phn. & 729 & \textbf{3.72} & \textbf{3.47$\pm$0.34} \\
\rowcolor[HTML]{EFEFEF} 
FunCodec w/ phn. & \textbf{250} & 6.73 & 2.84$\pm$0.35 \\ \midrule\midrule
\textbf{TTS} &  &  &  \\ \midrule
wav2vec2.0 & 729 & 2.21 & \textbf{4.42$\pm$0.11} \\
\rowcolor[HTML]{EFEFEF} 
FunCodec & \textbf{250} & \textbf{2.11} & 4.41$\pm$0.11 \\ 
\midrule\midrule
\textbf{TTS-1h} &  &  &  \\ \midrule
wav2vec2.0 & 729 & 41.99 & {2.77$\pm$0.37} \\
\rowcolor[HTML]{EFEFEF} 
FunCodec & \textbf{250} & \textbf{8.87} & \textbf{2.77$\pm$0.28} \\ \bottomrule
\end{tabular}
\end{table}

Table \ref{tab:tts} presents the vocoder resynthesis and the TTS results. In this table, the TTS results were obtained by the phoneme-augmented (``w/ phn.") version of vocoders. We used Whisper~\cite{whisper} to evaluate the intelligibility and UTMOS~\cite{utmos} to measure naturalness in a objective manner.
It can thus be seen that providing aligned phonemes in vocoders decreases the WER while also increases the UTMOS score when reconstructed from wav2vec2.0 tokens, which is the reason we only trained the FunCodec vocoder with phonemes.
Although FunCodec exhibits worse WER in resynthesis, we found it somehow improved the WER compared to wav2vec2.0 in the TTS scenario while keeping almost the same UTMOS.
Note that both tokens outperform ground truth in terms of WER. The reason might be that there are more tokenization errors from the real waveforms, but fewer from synthesized ones, and the vocoder still learns the right pronunciation from most of the correctly generated tokens.
Both wav2vec2.0 and FunCodec resulted in a TTS model that performed much better than the official baseline (HuBERT + FastSpeech) in terms of UTMOS.

For versions trained on only 1 hour data, we found FunCodec exhibited more stable decoding compared to wav2vec2.0 though the latter obtained better resynthesis quality. This can be explained by the drastical difference between the two vocabulary sizes.
FunCodec's 1024 tokens made prediction process much simpler.
From Table \ref{tab:tts}, we conclude that the first layer of FunCodec is the better choice for this single-speaker TTS challenge, since it has a much lower bitrate, better WER and still highly competitive UTMOS.

\section{SVS Track: The DOVSinger System}
The SVS track requires participants to build a discrete token-based SVS on the Opencpop ~\cite{opencpop} dataset, with a favorably low bit-rate and high naturalness score.
Our best submission used DAC as the discrete tokens and a modified VALL-E~\cite{valle} system as the SVS pipeline. 

\subsection{Discrete Tokens}
For singing voice synthesis (SVS), discrete tokens can also be categorized into semantic tokens and acoustic tokens, but discrete tokens specifically designed and trained for SVS are not very common. Recent work Make-A-Voice~\cite{makeavoice} utilized HuBERT~\cite{hubert} as the semantic tokens and SoundStream~\cite{soundstream} as the acoustic tokens and designed a unit-based vocoder to generate speech and singing voices. However, solely for SVS, we do not need to employ such complex network architectures. Through experimentation, we found that the Descript Audio Codec (DAC)~\cite{dac} can faithfully restore the original audio in  the Opencpop dataset.

DAC is a universal neural audio compression algorithm renowned for its high-fidelity performance, achieving compression of 44.1 kHz audio into tokens. The sampling rate of Opencpop is also 44.1kHz, avoiding information loss during downsampling. In contrast, other acoustic token models like SoundStream~\cite{soundstream}, tailored for audio with a sampling rate of 24kHz, necessitate downsampling, leading to potential information loss.
Moreover, DAC endeavors to compress audio across all domains and benefits from training on datasets abundant in music, such as the MUSDB dataset~\cite{Rafii2017MUSDB18A}, consequently yielding outstanding reconstruction results for singing voices.

In consideration of the excessive information load and significantly high bitrate carried by the 18-layer DAC, we ultimately opted to utilize the the initial layer of DAC features  as discrete tokens to facilitate information transmission.

\subsection{Model Architecture}

\subsubsection{Acoustic Model}
\begin{figure}
    \centering
    \includegraphics[width=0.9\linewidth]{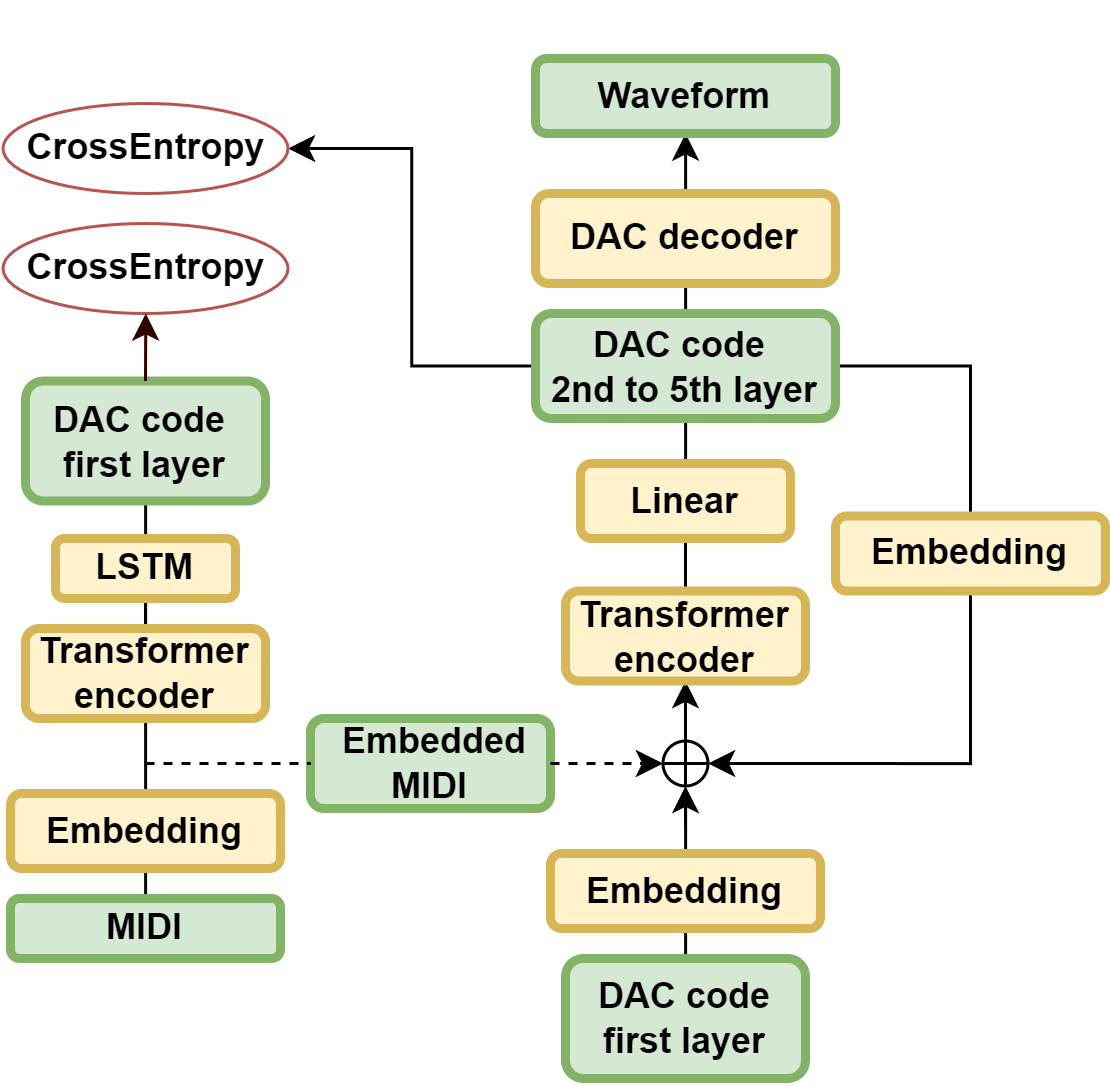}
    \caption{Architecture of the acoustic model (left) and vocoder (right) in the SVS track. }
    \label{fig:svs}
\end{figure}

The acoustic model has been refined through adjustments grounded in VALL-E~\cite{valle}'s framework. The model architecture is visualized in Fig.\ref{fig:svs}.  Within the existing framework, VALL-E employs a decoder-only neural network architecture to decode input information directly into discrete tokens. Specifically, it utilizes a single-layer autoregressive (AR) structure to decode the first-layer tokens, and employs a seven-layer non-autoregressive (NAR) structure to decode the second to eighth layer. 

In the AR architecture, VALL-E performs prediction in the temporal axis. However, in the context of the SVS task, as duration is explicitly provided in the MIDI input, there is no requirement for supplementary mechanisms to predict duration. Therefore, after augmenting the phone, pitch, slur, and other information with phone duration from the MIDI, we directly employed the NAR structure to predict the first-layer DAC codes. 

After experimentation, we observed that the stability of the reconstructed waveform was compromised when employing the NAR structure for predicting the first-layer codes directly. To facilitate the learning of contextual information from neighboring frames more effectively, we introduced an additional layer of LSTM following the transformer layer. The results demonstrated that the waveform reconstructed in this manner exhibited improved temporal stability and coherence. The training criterion is the cross entropy loss of DAC's first layer prediction.

\subsubsection{Discrete Unit-based Vocoder}

The discrete code information from the first layer of the DAC obtained after the acoustic model is combined with the MIDI information as inputs to the vocoder. 

The primary architecture of the vocoder involves the continuation of inferring subsequent DAC codes using input information, followed by waveform reconstruction using a DAC decoder. During the inference of each layer of code, additional information from preceding layers is also incorporated.To ensure both the speed and quality of vocoder generation, we adopted a fully NAR structure to further infer four layers of DAC code, replacing the LSTM structure with a Linear one to enhance inference speed.The training criterion is the cross entropy loss of DAC's 2nd to 5th layer prediction.

\subsection{Data Preparation and Training}

\subsubsection{Data Preparation}
The provided dataset is Opencpop~\cite{opencpop}, consisting of 100 unique Mandarin songs, which were recorded by a professional female singer. Furthermore, we leveraged the M4Singer dataset during training to augment the model's capacity for robust fitting. It is essential to underscore that the incorporation of supplementary datasets mandates the inclusion of speaker attributes within the MIDI input. Prior to training, we performed DAC extraction on the entire training corpus to obtain discrete codes. Additionally, we converted phone durations into frame counts using DAC's 1/86-second frame shift and temporally extended other MIDI attributes to match the frame count length, enabling direct integration into the NAR model.

\subsubsection{Training}
We trained the acoustic model for 160 epochs using the Adam optimizer at a base learning rate of 0.05.
The transformer encoder contains 6 layers, each with 512 attention dimensions and 8 heads.

\subsection{Experimental Results}
\begin{table}[!h]
\centering
\caption{SVS results in the SVS track. All data is sourced from the leaderboard. However, the leaderboard did not provide the bitrate and MOS of baseline system.}
\label{tab:svs}
\begin{tabular}{@{}lccc@{}}
\toprule
 & \begin{tabular}[c]{@{}c@{}}\textbf{Bitrate}$\downarrow$\\ \footnotesize{(bps)}\end{tabular} & \begin{tabular}[c]{@{}c@{}}\textbf{Log F0 RMSE}$\downarrow$\end{tabular} & \textbf{MOS}$\uparrow$ \\ \midrule
baseline & - & 0.19 &  - \\ 
DOVSinger & 725.9 & 0.19 & 2.63 \\ \bottomrule
\end{tabular}
\end{table}

Table \ref{tab:svs} presents the SVS results and all the data is sourced from the leaderboard. From the table we could found that, at the expense of reducing bitrate while simultaneously ensuring accuracy in pitch, DOVSinger sacrifices perceptual quality. One reason is the utilization of only five layers of DAC code for reconstruction, resulting in noticeable hoarseness in the final vocal output. Another factor is the missed opportunity for improvement by directly employing the DAC decoder for reconstruction, since employing a vocoder for decoding, while incorporating the loss of Mel-spectrogram reconstruction during training, could enhance the quality of the generated vocals.
\section{ASR Track}

\subsection{Discrete Tokens for ASR}
Speech discrete tokens for ASR can be broadly categorized into semantic tokens and acoustic tokens based on linguistic information or acoustic details of the speech.
Specifically, models including HuBERT~\cite{hubert} and WavLM~\cite{wavlm} generate semantic tokens, which are trained for discrimination tasks or masking predictions while audio neural codec models like Encodec~\cite{encodec} yield acoustic tokens aimed at speech reconstruction.

\subsection{ASR with Discretized Input}
\begin{figure}[!h]
    \centering
    \includegraphics[width=0.9\linewidth]{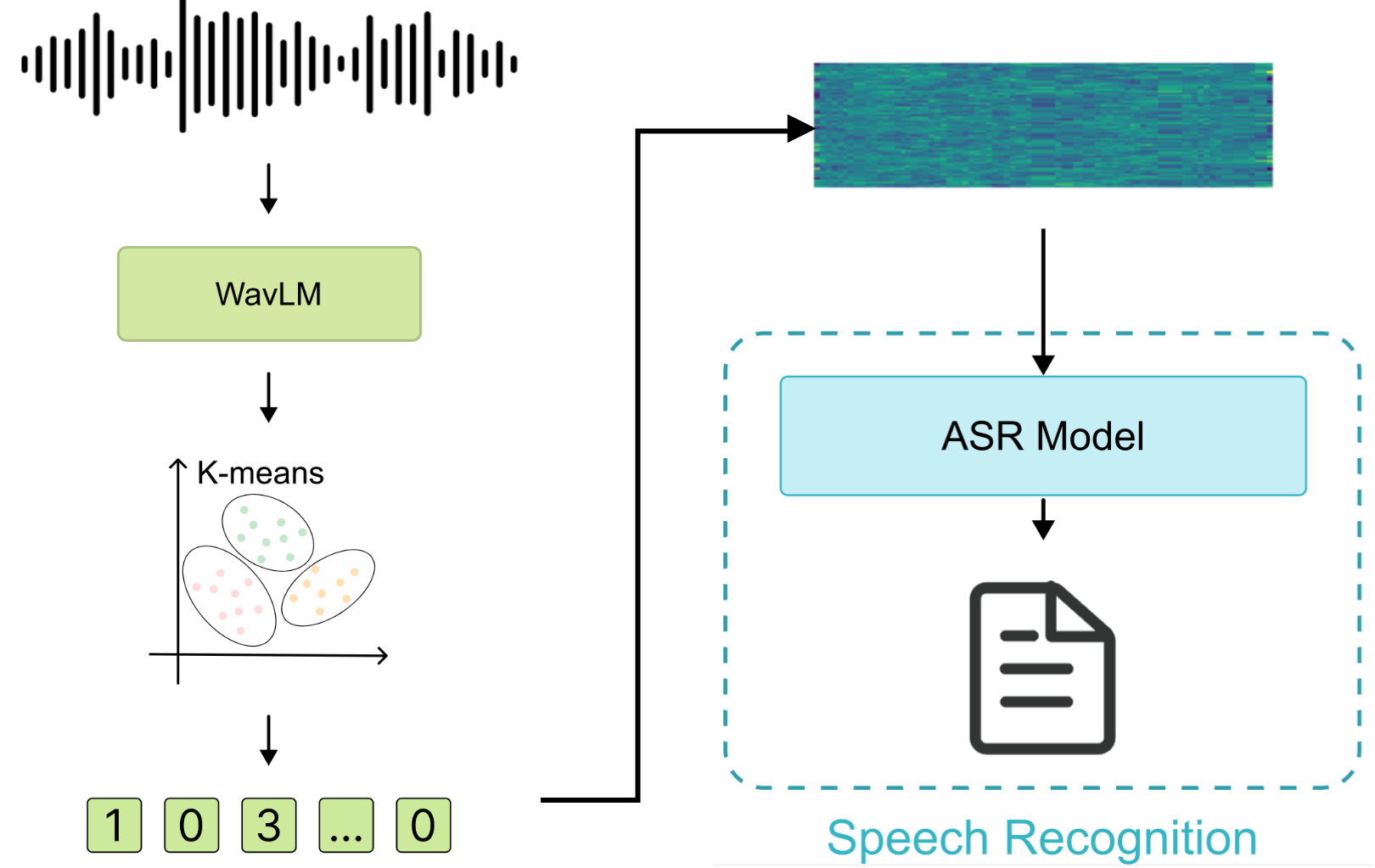}
    \caption{Illustration of the pipeline for speech discrete tokens in the ASR track.}
    \label{fig:asr}
\end{figure}

\subsubsection{Speech Discretization}
In the ASR track, we used WavLM-large to extract features, and then applied k-means with 2000 classes to obtain discrete tokens.
Following~\cite{yang2024towards}, we performed the data augmentation techniques for discretized inputs to enhance robustness for discrete token representations. 
Along with their corresponding texts, we leveraged Zipformer~\cite{zipformer} as the encoder and RNN-T loss~\cite{rnnt} for optimization to train an end-to-end ASR model.
These tokens were projected to 80 dimensions through a linear embedding layer.
Then, these features were duplicated twice frame by frame to a uniform 100 Hz rate before feeding into the ASR model.

\subsubsection{Model Architecture}
The neural Transducer architecture was adopted for ASR. Pruned RNN-T loss~\cite{prunedrnnt} was used as the training objective function, implemented within the k2~\cite{k2} framework\footnote{https://github.com/k2-fsa/k2}. The encoder employed a 6-stack Zipformer~\cite{zipformer} with downsampling factors of (1,2,4,8,4,2). The label decoder employed a stateless decoder~\cite{stateless}, which consists of an embedding layer followed by a 512-dim Conv1D layer.
A convolution subsampling module with a stride of 2 was placed to reduce the frame rate to 50 Hz before being fed into the encoder. Overall, the model had 65.5M parameters.

\subsubsection{Training}
We trained the acoustic model for 100 epochs using the ScaledAdam~\cite{zipformer} optimizer at a base learning rate of 0.045. We utilized 500-class Byte Pair Encoding (BPE)~\cite{bpe} word pieces as the classification units. 

\subsection{Experimental Results}
The ASR track used LibriSpeech~\cite{librispeech} and ML-SUPERB~\cite{ml-superb} corpora to evaluate.
Performance metrics include character error rate (CER) and bitrate.
As illustrated in Table \ref{tab:asr}, our system achieved a relative CER reduction of up to 13.0\% compared to the baseline, albeit at a higher bitrate of 550 bps.

\begin{table}[!h]
\centering
\caption{CER and bitrate of our system in the ASR track.}
\label{tab:asr}
\begin{tabular}{@{}lcc@{}}
\toprule
 & \begin{tabular}[c]{@{}c@{}}\textbf{Bitrate (bps)}$\downarrow$ \end{tabular} & \begin{tabular}[c]{@{}c@{}}\textbf{EN\_LibriSpeech CER (\%)}$\downarrow$\end{tabular} \\ \midrule
baseline  & 335.86 & 2.31    \\
Zipformer & 550.00 & \textbf{2.01}  \\ \bottomrule
\end{tabular}
\end{table}

\vspace{-0.15in}
\section{Conclusion}
We present the systems for TTS, ASR and SVS track in the discrete speech unit challenge.
Detailed model designs and choices of discrete tokens are illustrated.
Specifically, we use FunCodec and an improved VQTTS model for TTS to achieve a low bitrate yet highly competitive TTS system that ranks the first.
DAC and VALL-E are adopted in the SVS track, while k-means clusters of WavLM and Zipformer are utilized in ASR track.
We hope our experimental findings could promote better understanding and utilization of the discrete speech tokens among the speech community.
\newpage
\bibliographystyle{IEEEtran}
\bibliography{mybib}

\end{document}